\documentclass[10pt,journal]{IEEEtran}

\usepackage{color}
\usepackage{amsmath,amssymb,amsfonts}
\usepackage{graphicx}

\usepackage{hyperref}
\graphicspath{{./Figures/}{./}}
\newcommand{\R}{{\mathbb R}}

\usepackage{mathtools}

\begin{document}

\title{Probing high order dependencies with information theory}

\author{\IEEEauthorblockN{C. Granero-Belinch\'on\IEEEauthorrefmark{1}}, \IEEEauthorblockN{S.G. Roux\IEEEauthorrefmark{1}}, \IEEEauthorblockN{P. Abry (IEEE Fellow)\IEEEauthorrefmark{1}}, \IEEEauthorblockN{N.B. Garnier\IEEEauthorrefmark{1}\\}\IEEEauthorblockA{\IEEEauthorrefmark{1}Univ Lyon, Ens de Lyon, Univ Claude Bernard, CNRS, Laboratoire de Physique, F-69342 Lyon, France.\\}\thanks{This work was supported by the LABEX iMUST (ANR-10-LABX-0064) of Universit\'e de Lyon, within the program "Investissements d'Avenir" (ANR-11- IDEX-0007) operated by the French National Research Agency (ANR).}%
}

\maketitle

\begin{abstract}
Information theoretic measures (entropies, entropy rates, mutual information) are nowadays commonly used in statistical signal processing for real-world data analysis. 
The present work proposes the use of Auto Mutual Information (Mutual Information between subsets of the same signal) and entropy rate as powerful tools to assess refined dependencies of any order in signal temporal dynamics. 
Notably, it is shown how two-point Auto Mutual Information and entropy rate
unveil information conveyed by higher order statistic and thus capture details of temporal dynamics that are overlooked by the (two-point) correlation function.
Statistical performance of relevant estimators for Auto Mutual Information and entropy rate are studied numerically, by means of Monte Carlo simulations, as functions of sample size, dependence structures and hyper parameters that enter their definition. 
Further, it is shown how Auto Mutual Information permits to discriminate between several different non Gaussian processes, having exactly the same marginal distribution and covariance function. 
Assessing higher order statistics via multipoint Auto Mutual Information is also shown to unveil the global dependence structure fo these processes, indicating that one of the non Gaussian actually has temporal dynamics that ressembles that of a Gaussian process with same covariance while the other does not. 
\end{abstract}

\IEEEpeerreviewmaketitle

\section{Introduction}
\label{sec:intro}

\noindent {\bf Context.} To characterize a (stationary) process $X(t)$, evolving along a given dimension $t$, \textit{e.g.}, time, the precise assessment of both its static properties (marginal distribution or one-point statistics) and temporal dynamics (joint distributions or multi-point statistics) is crucial. 

Correlation and spectral estimates,  providing practitioners with valuable insight into temporal dynamics are key tools in statistics and data analysis 
and have been widely used in many different fields.
In neurosciences, the analysis of correlation between neuron firings~\cite{Moore1970,Cohen2011} or between stimulus and brain activity~\cite{Gray1996,Miller2009} is essential.
In fluid mechanics, Kolmogorov's empirical theory of turbulence~\cite{Kolmogorov1941a} characterizes the multiscale distribution of energy with the power spectrum of the velocity field~\cite{F.Anselmet1984}. 
The analysis of correlations also supports useful descriptions in terms of complex networks for, e.g., ecology~\cite{Zhang2011}, climatology~\cite{Yamasaki2008} or traffic~\cite{Ou2007}.
 
It is however well known that, while the correlation function, or equivalently the power spectral density, exhaust the description of joint statistics and hence of temporal dynamics for  jointly Gaussian processes, they provide characterizations of temporal dynamics for general processes, that are limited in at least two respects: not only they are restricted to two-point joint statistics but also to a specific measure of these two point-statistics. 
In general, the full analysis of temporal dynamics require the use of multi-point, or higher order, joint statistics. 
The goal of the present work is to address such an issue using information theoretic measures.\\

\noindent {\bf Related work.}  To assess dependencies beyond two-point correlations, multipoint correlation functions were defined. 
For instance, three-point correlation functions have been used in cosmology, where they were related to cosmological parameters~\cite{Takada2002,Slepian2015}. 
Alternatively, higher order-spectra, that combine Fourier transforms at several different frequencies and therefore takes into account higher order statistics, have also been used to generalize the power spectrum beyond second-order statistics~\cite{Nikias1993,Lacoume1997}. 
For instance, bi-spectrum and tri-spectrum were analyzed in depth in e.g., \cite{Collis1998,Chandran1995} and used in applications as different as cosmology~\cite{Matsubara2010, Trivedi2012}, or medicine~\cite{Ning1990,Husar1997} and other fields~\cite{Astola2008, Courtney2010}.
While of popular use in applications, higher-order correlation functions and spectra suffers from major caveats, mostly related to estimation issues. 
Indeed, the assessment of order $m$-statistics with higher-order correlation functions and spectra requires the estimation of $m$-time point statistics, which, for a given fixed sample size, can only yield decreasing estimation performance when order $m$ is increased. 

Alternatively, Shannon's information theory, and the corresponding information theoretic measures, entropy, entropy rate and mutual information, can be considered as an efficient way to characterize high order statistics \cite{Shannon1948,Shannon1949,Cover2006}. 
Applications of entropy rate were yet reported in a wide range of domains, such as biology~\cite{Herzel1994}, dynamical systems~\cite{Lesne2009}, fluid turbulence~\cite{Granero16} or analysis of languages~\cite{Takahira2016}.
Mutual information is also very popular, especially in computer sciences, e.g., in machine learning~\cite{Tourassi2001}, and other very different domains~\cite{Granger1994,Michel1996,Darbellay1997,Pham2004,Duncan2008, Malladi2018}.
Auto-Mutual Information (AMI) ---~sometimes called Information Storage~\cite{Xiong2017}~---, defined as the mutual information between the analyzed process and a delayed version of itself, has been used in different domains~\cite{Bernhard1998,Albers2012, Albers2012a,C.Granero-Belinchon2017}.
Information theory additionally provides tools to explore causality relationships between processes~\cite{Amblard2011,Amblard2013}. However, within this paper we are mainly interested in the characterization of their dependence structures.

Entropy rate and AMI have been envisaged as tools to describe temporal dynamics in time series beyond second-order statistics, or even at second-order statistics, in a richer way than correlation or Fourier spectrum, see~ \cite{Gaspard1993,Bernhard1998,Albers2012a,Granero16}.

\noindent {\bf Goals, contributions and outline.}  The main goals of the article are
i) to show the ability of AMI and entropy rate to describe high order dependences, even when considering only two-point interactions. Even though this property of information theory tools is well known, we are able to show it by analyzing processes with identical correlation and marginal distribution but different high order dependences. AMI and entropy rate appear as straightforward generalizations of correlation function and power spectrum when considering non-Gaussian processes analysis.
ii) to show the effect of analysing interactions between more than two points, combining Takens embedding~\cite{Takens1981} with AMI. We characterize the complexity of the dependence structure of a process by measuring AMI for embedding dimension higher than 2. More precisely, we analyze the differences between power law decay and exponential decay dependence structures. 
 
To demonstrate the ability of information theory to describe high order statistics, we illustrate the use of entropy rate and auto-mutual information to characterize non Gaussian processes across scales.
As an example, we consider two synthetic log-normal processes with different dependence structures but with identical correlation function and identical marginal statistics.
In section~\ref{sec:InfTheo}, we introduce and define the theoretical information theory measures and our estimators. In section \ref{sec:ami},  we interpret the information theory tools used along the article,  we present the processes with identical correlation function and one-point statistics but different dependence structure that we study and finally we illustrate how AMI and entropy rate allow to distinguish between them.  
In section~\ref{ssec:Procedure}, we report the convergence and the performance of our estimators of AMI and entropy rate and their ability to measure high order dependencies. We present their evolution across scales and emphasized that, contrary to classical correlation analysis, they allow a fine characterization of non Gaussian processes.

%%%%%%%%%%%%%%%%%%%%%%%%%%%%%%%%%%%%%%%%%%%%  
\section{Entropy, entropy rate and auto mutual information}
\label{sec:InfTheo}

\subsection{Definitions}

To analyze the temporal dynamics of univariate stationary processes $X=\left\lbrace x_t \right\rbrace_{t\in\mathbb{R}}$, it has been proposed in \cite{Takens1981} to rely on the time delay-embedding procedure. It amounts to constructing the $m$-dimensional process $X^{(m,\tau)}$, whose elements are $m$-dimensional vectors: 
\begin{equation}
\textbf{x}_t^{(m,\tau)}=\left(x_t, \,x_{t-\tau}, \, \cdots, \, x_{t-(m-1)\tau}\right)\,, 
\end{equation}
where the delay $\tau$ implicitly defines an analysis scale and where the embbeding dimension $m$ controls the order of the statistics of $X$ which are analyzed. 
The information theoretic measures defined below aim to characterize temporal dynamics via the analysis of the $m$-time point joint statistical distributions $p(\textbf{x}^{(m,\tau)}_{t})$. 

Within this article, we study ergodic stationary processes and then $p(\textbf{x}^{(m,\tau)}_{t})$ depend on $m$ and $\tau$ but not on $t$. Consequently $p(\textbf{x}^{(m,\tau)}_{t})=p(X^{(m,\tau)})$. 

{\bf Shannon entropy}, is defined as a functional of $p(\textbf{x}^{(m,\tau)}_{t})$ \cite{Shannon1948}:
\begin{eqnarray*}
H(\textbf{x}_t^{(m,\tau)})&=&- \int_{\R^m} p(\textbf{x}_t^{(m,\tau)})\ln
p(\textbf{x}_t^{(m,\tau)}) {\rm d}\textbf{x}_t^{(m,\tau)} \\
&=&H^{(m,\tau)}(X)
\end{eqnarray*}
By nature, it depends on the full joint distribution and therefore on statistics of any order.
However, for stationary processes it does not depend on the mean value of the marginal distribution.
For $m=1$, Shannon entropy does not relate to temporal dynamics but to one-point statistics and thus obviously does not depend on $\tau$ and is hence denoted $H(X)$. 

{\bf Entropy rate} of order $m$, $h^{(m,\tau)}$, quantify variations of Shannon entropy between two successive time-embedded versions of a stationary process~\cite{Shannon1948,Cover2006}: 
\begin{align}
h^{(m,\tau)}(x_t) &= H(\textbf{x}_t^{(m+1,\tau)}) - H(\textbf{x}_{t-\tau}^{(m,\tau)}) \nonumber\\
&=  H^{(m+1,\tau)}(X) - H^{(m,\tau)}(X) \nonumber \\ 
&= h^{(m,\tau)}(X). 
\label{eq:def:h:diff}
\end{align}
Entropy rate $h^{(m,\tau)}$, like $H^{(m,\tau)}$, probes joint statistics of any order, but in addition quantifies how much information is gained in the analysis of temporal dynamics by increasing by one the order of the embedding.  

 {\bf Mutual information} $I(\textbf{x}_t^{(m,\tau)},\textbf{y}_{t'}^{(p,\tau)})$ quantifies the amount of information shared by two processes $X$ and $Y$, via their embbeded versions $\textbf{x}_t^{(m,\tau)} $ and $\textbf{y}_{t'}^{(p,\tau)}$~\cite{Shannon1948,Cover2006}:
\begin{align}
I(\textbf{x}_t^{(m,\tau)}, &\textbf{y}_{t'}^{(p,\tau)})=\int_{\R^{m+p}} p\left(\textbf{x}_t^{(m,\tau)},\textbf{y}_{t'}^{(p,\tau)}\right)  \nonumber  \\
&\ln \left(\frac{p\left(\textbf{x}_t^{(m,\tau)},\textbf{y}_{t'}^{(p,\tau)}\right)}{p(\textbf{x}_t^{(m,\tau)})p(\textbf{y}_{t'}^{(p,\tau)})}
\right) {\rm d}\textbf{y}_{t'}^{(p,\tau)} {\rm d}\textbf{x}_t^{(m,\tau)}
\label{eq:def:MI}
\end{align}
When $X$ and $Y $ are jointly stationary processes, $I(\textbf{x}_t^{(m,\tau)},\textbf{y}_{t'}^{(p,\tau)})=I(X_t^{(m,\tau)},Y_{t'}^{(p,\tau)})$ obviously depends on $t'-t$ only.

Mutual information can be read as the Kullback-Leibler divergence between the joint probability $p(\textbf{x}_t^{(m,\tau)},\textbf{y}_{t'}^{(p,\tau)})$ and its counterpart under independence hypothesis $p(\textbf{x}_t^{(m,\tau)})p(\textbf{y}_{t'}^{(p,\tau)})$. 

To analyze the higher statistical order temporal dynamics (also referred to as nonlinear temporal dynamics) of a single stationary process $X$, one can use auto-mutual information~\cite{Cover2006, Xiong2017}, defined as the mutual information between $\textbf{x}_t^{(m,\tau)} $ and $\textbf{x}_{t+p\tau}^{(p,\tau)} $:
\begin{equation}
I^{(m,p,\tau)}(X)=  I(\textbf{x}_{t}^{(m, \tau)},\textbf{x}_{t+p\tau}^{(p, \tau)}) \,.
\label{eq:def:AMI}
\end{equation}
Because the concatenation of $\textbf{x}_{t+p\tau}^{(p, \tau)}$ and $\textbf{x}_{t}^{(m, \tau)}$ corresponds exactly to the $(m+p)$-dimensional time-embedded vector $\textbf{x}_{t+p\tau}^{(m+p)}$:
\begin{equation*}
\left(\textbf{x}_t^{(m,\tau)},\textbf{x}^{(p,\tau)}_{t+p\tau}\right)=\textbf{x}_{t+p\tau}^{(m+p,\tau)}, 
\end{equation*}
$I^{(m,p,\tau)}(X)$ quantifies the shared information between the past $m$-point dynamics and  the future $p$-point dynamics. 
Auto mutual information quantifies the deviation from independence between the two arguments, beyond the simpler decorrelation.

These information theoretic quantities all depend on the chosen time lag $\tau$. 
When $\tau\to 0$, dependences between two consecutive coordinates of the embedding vector increase, therefore inducing an increase in  $I^{(m,1,\tau)}(X)$ and a decrease in $h^{(m,\tau)}(X)$.
Conversely, when $\tau\to +\infty$, dependencies decrease to $0$, therefore inducing an increase of $h^{(m,\tau)}(X)$ up to the Shannon entropy $H(X)$ of the process, its largest value, and a decrease of $I^{(m,1,\tau)}(X) \rightarrow 0$.

Furthermore, entropy rate can then be expressed as \cite{Granero16}:
\begin{equation}
h^{(m,\tau)}(X) =  H(X) - I^{(m,1,\tau)}(X) \,,
\label{eq:def:h:MI}
\end{equation}
which shows that there are two different contributions to the entropy rate: 
First, $H(X)$, which only depends on the one-point distribution and hence is a static property~;
second, $I^{(m,1,\tau)}(X)$ which gathers all information conveyed by linear and non linear temporal dynamics, irrespective of the variance of the distribution~\cite{Granero16}.

\subsection{Gaussian process}

When $X$ is a stationary jointly Gaussian process, hence fully defined by its variance $\sigma_{x}^2$ and normalized correlation function $c_{X}(\tau)$, analytical expressions for entropy, entropy rate and auto mutual information are available~\cite{Zografos2005}:
\begin{align}
H(X) = & \frac{1}{2}\ln(2 \pi e \sigma_{x}^{2}) \label{eq:Hgauss}\\
I^{(m,p,\tau)}(X)  = & \frac{1}{2}\ln\left(\frac{|\Sigma^{(m)}||\Sigma^{(p)}|}{|\Sigma^{(m+p)}|}\right) \label{eq:gauss} \\
h^{(m,\tau)}(X) = & \frac{1}{2}\ln(2 \pi e \sigma_{x}^{2}) - \frac{1}{2}\ln\left(\frac{|\Sigma^{(m)}|}{|\Sigma^{(m+1)}|}\right)  
\end{align}
with $\Sigma^{(m)}$ the $m$-dimensional correlation matrix, and $||$ denoting the determinant. By definition,  $\Sigma^{(1)}=1$. 

%%%%%%%%%%%%%%%%%%%%%%%%%%%%%%%%%%
\subsection{Estimation procedures}

To estimate these information theoretic quantities, two main categories of non-parametric procedures are available, based on either 
kernel density estimations (KDE) or
nearest neighbor search (NNS). 
While the former essentially counts the number $k(\epsilon)$ of the observed samples in interval of a priori set size $\epsilon$, the latter rather measure the size $\epsilon(k) $ of box that contain an a priori set number of $k$ samples. 

Because this second strategy has been documented as providing low bias estimates \cite{Grassberger2004,Gao2016}, we chose $k$-nearest neighbors procedures to estimate the entropy $H(X)$~\cite{L.Kozachenko1987,Singh2003, Amblard2008}:

\begin{equation}\label{eq:koza}
\widehat{H}(X) = -\psi(k) + \psi(N) + \frac{d}{N}\sum_i^N \log(\epsilon_i(k))
\end{equation}

\noindent with $\psi$ the Digamma function, $d$ the dimension of $X$, $N$ the number of samples of $X$, and using the Chebyshev distance ($L_{\infty}$ metric).

To compute the mutual information, we use the $k$-nearest neighbors procedure presented by Kraskov, et al.~\cite{Grassberger2004}: 
\begin{equation}
\widehat{I}(X,Y)=
\psi(k)+\psi(N)-\left\langle\psi(n_{x})+\psi(n_{y})\right\rangle
\end{equation}

The procedure chooses $k$ for the joint distribution $Z(X,Y)$ and hence obtains the distance $\epsilon$ to the $k$th nearest neighbour. 
Then using $\epsilon$, it calculates the number of neighbours $n_x+1$ lying on a segment $\textbf{x}= x \pm \frac{\epsilon}{2}$ of the X dimension, \textit{i.e.} lying on 
$-\infty\leq Y \leq \infty$, $x-\epsilon/2 \leq X \leq x+\epsilon/2$. Consequently, $\epsilon/2$ is the distance to the $n_x+1$st neighbour of $x$. The procedure uses the same development for $Y$ and obtains that $\epsilon/2$ is the distance to the $n_y+1$st neighbour of $y$. 

Because benefits against a direct use of eq.(\ref{eq:def:h:diff}) were documented in \cite{Granero2017Gretsi},  $h^{(m,\tau)}(X)$ is estimated plugging the above estimates into eq.(\ref{eq:def:h:MI}). 

For an stationary process $X$, $p(\textbf{x}^{(m,\tau)}_{t})$ does not depend on $t$, and consequently, we can exploit this property to estimate the information theoretic quantities of $X$ by averaging on all the temporal samples of the process.  

To perform an analysis across scales by varying the parameter $\tau$ we use an adapted Theiler prescription~\cite{Theiler1986} to eliminate the contribution of
dependencies from scales smaller than $\tau$.

\section{Temporal dynamics and dependence structure}
\label{sec:ami}

The aim of this section is to try to characterize what part of the statistics of a given stationary process $Y$ contribute to the values taken by information theoretic measures $H$, $h^{(m,\tau)}$ and $I^{(m,p,\tau)}$. 
To that end, we conduct a formal study of the specific yet very instructive case where processes $Y$ with correlation function $c_Y(\tau)$ and marginal distributions $p_Y$ are obtained as point bijective transformation $F$, $Y=F(X)$ of an underlying Gaussian process $X$ with correlation function $c_X$, designed to reach the targeted covariance $c_Y$. Processes $Y$ constructed in this way depend on $F$ and $c_X(\tau)$ \cite{Helgason2011a,Helgason2011}.

\subsection{Analytical calculations and interpretations}
\label{sec:ami:a}
Given any bijective transformation $F$ that maps a Gaussian signal $X$ into a signal $Y=F(X)$, the information theoretic quantities of $Y$ can be computed as:
\begin{align}
H(Y) &= H(X) + \langle \ln F \rangle_X  \nonumber \\
&= \frac{1}{2}\ln(2 \pi e \sigma_{X}^{2}) + \langle \ln F \rangle_X \label{eq:generic:H} \\
I^{(m,p,\tau)}(Y)  &= I^{(m,p,\tau)}(X) \nonumber \\
&= \frac{1}{2}\ln\left(\frac{|\Sigma_X^{(m)}| |\Sigma_X^{(p)}|}{|\Sigma_X^{(m+p)}|}\right) \label{eq:generic:MI}\\
h^{(m,\tau)}(Y) = &  \frac{1}{2}\ln(2 \pi e \sigma_{X}^{2}) + \langle \ln F \rangle_X -  \frac{1}{2}\ln\left(\frac{|\Sigma_X^{(m)}|}{|\Sigma_X^{(m+1)}|}\right)  \label{eq:generic:h}
\end{align}
with $\langle \ln F \rangle_X =  \int  p(\textbf{x}_t^{(1)}) \ln F(x) dx$, hence a constant, $\sigma_{X}^2$ and $\Sigma_X^{(m)}$ are the variance and the $m$-dimensional covariance matrix of the underlying Gaussian process $X$. 

These analytical calculations permit to yield understanding of general validity regarding the nature of the information conveyed by each of these three information theoretic quantities: \\
i) Entropy $H$ is only related to the marginal distribution of the process $p_Y$, hence is quantifying static properties only.\\
ii) Entropy rate $h^{(m,\tau)}$ mixes up marginal distribution features (variance and pointwise transform $F$), hence static properties with dynamical properties and dependencies of any order, as indeed $\Sigma_X^{(m)}$ depends jointly on $c_Y$ and on $F$, hence on the joint distribution of $\textbf{y}_t^{(m,\tau)}$.
The fact that $h^{(m,\tau)}$ gather both static and dynamics aspects of time series likely explains why it has been observed in many applications to be a often discriminant feature, compared to other features \cite{C.Granero-Belinchon2017}. 
However, the entanglement of static and dynamic properties may also be considered a drawback for the analysis, characterization and understanding of data properties. \\
iii) Interestingly, auto mutual information $I^{(m,p,\tau)}$ performs a characterization mainly related to the dependence structure of $Y$, as indeed $\Sigma_X^{(m)}$ depends on the joint distribution of $\textbf{y}_t^{(m,\tau)}$. It therefore quantifies temporal dynamics, irrespective of  static properties. Thus, AMI is the sole measure that focuses exclusively on higher order dynamics and on the joint dependence structure, beyond correlation. 

Often in the scientific literature, $h^{(m,\tau)}$ and $I^{(m,p,\tau)}$ are referred to as \emph{nonlinear} features, to emphasize that they characterize temporal dynamics beyond the mere use of the correlation functions or power spectral densities. 
This is indeed correct. 
Even when two-point statistics are analyzed, i.e., $m=p=1$, $I^{(1,1,\tau)}$ reflects dependencies, including and encompassing correlations as the whole joint two-point distribution function is involved. 
Nevertheless,  $I^{(m,p,\tau)}$ remains mostly driven by the correlation function $c_Y$, and hence by \emph{linear} effects. 
This is also the case for $h^{(m,\tau)}$, driven mostly both by the static variance $\sigma^2_Y$ and dynamical correlation function $c_Y$. 

\subsection{Process definition}

\subsubsection{Analytical derivations}

To further illustrate that indeed $I^{(m,p,\tau)}$ and $h^{(m,\tau)}$ may capture differences in any-order dependencies beyond correlations, let us study the cases where two stationary processes $Y_{F_1}$ and $Y_{F_2}$ have exact same marginal distribution $p_Y$ and exact same covariance $c_Y$, but are yet obtained from two different point transformations $F_1$ and $F_2$ ---~applied to Gaussian processes $X_1$ and $X_2$~--- and hence have different joint distributions and hence different dependence structures.

Let denote $F_Y$ the cumulative distribution function associated to the marginal distribution $p_Y$.
A natural candidate to obtain a process with marginal $p_Y$ is to apply the point transform
\begin{equation}
\label{eq:F1}
F_1(x) =  F^{-1}_Y \circ \Phi(x), 
\end{equation}
to a Gaussian process $X_1$, where $F_Y$ is the cumulative distribution function of the marginal distribution $p_Y$ and $\Phi(x)$ the cumulative distribution function of a centered standard Gaussian variable, and where $X_1$ is a  jointly Gaussian process with covariance function $c_{X_1}(\tau)$ to reach the targeted $c_{Y}(\tau)$ \cite{Helgason2011a,Helgason2011}.

However, many other transforms can yield processes $Y$ with the exact same marginal distribution $p_Y$ and the exact same covariance $c_Y$.
In \cite{Helgason2011}, it was for example proposed to use 
\begin{equation}
\label{eq:F2}
F_2(x)= F^{-1}_Y \circ \left(2 (\Phi(|x|)-1/2) \right)\,, 
\end{equation}
on a jointly Gaussian process  $X_2$.
This thus yields a radically different joint dependence structure for $Y_{F_2}$, despite having the same prescribed correlation function as $Y_{F_1}$.
As a consequence, $Y_{F_1}$ and $Y_{F_2}$ have different AMI functions $I^{(m,p,\tau)}(Y_{F_1})  \neq I^{(m,p,\tau)}(Y_{F_2})$, and entropy rates, and hence different joint dependence structure and thus temporal dynamics. 
Conversely, their entropies are identical $H(Y_{F_1}) = H(Y_{F_2}) $. This is not surprising , as the entropy is completely defined by the one-point distribution.

\begin{figure*}%[tbh]
\begin{center}
\includegraphics[width=0.75\linewidth]{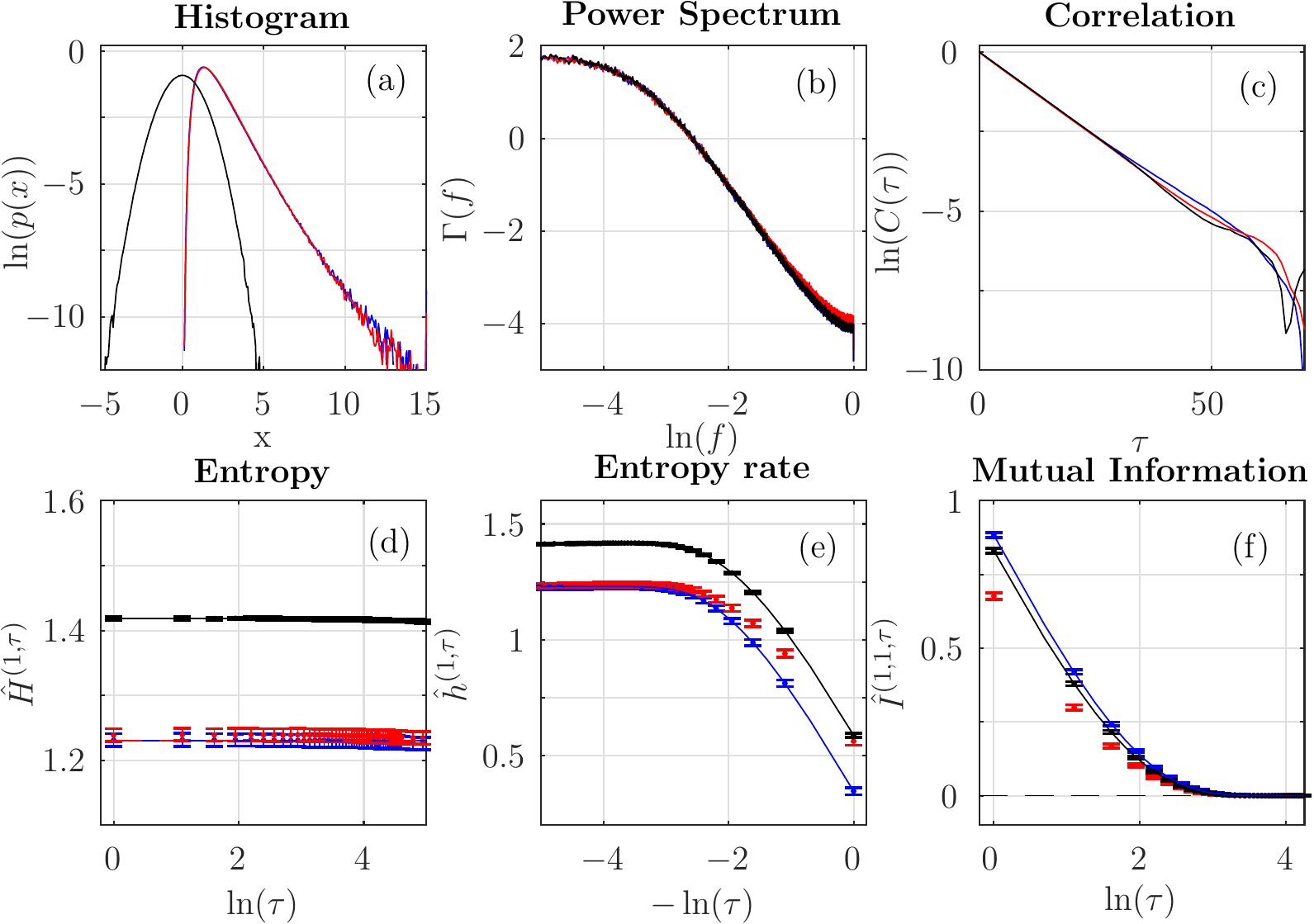}
\caption{{\bf  Short-range dependencies. } Classical analysis: histogram (a), power spectrum (b) and correlation function (c). Information theory analysis for $m=1$, $p=1$: entropy (d), entropy 
rate (e) and auto mutual information (f). 
The color indicates the process: blue for standard log-normal ($Y_{F_1}$), red for $Y_{F_2}$ log-normal and black for Gaussian ($Y_{\mathbb I}$).
The continuous lines in (e,d,f) correspond to analytical expressions. 
 We used $N=2^{16}$ and $k=5$.}
\label{fig:results:expp9}
\end{center}
\end{figure*}

\subsubsection{The log-normal example}
\label{sec:lognormal:theory}

To quantify the qualitative analysis proposed in Section~\ref{sec:ami:a} , let us consider stationary processes $Y_{F_1}$ and $Y_{F_2}$ having the same log-normal  marginal distribution $p_Y$,  chosen as an example of distributions that may complicate estimation because of its being a \emph{frontier} between distributions with all moments finite and heavy-tail distributions. 

For process $Y_{F_1}$ obtained with the natural transform (\ref{eq:F1}), $F_1$ simplifies to $F_1(x)=\exp(x)$, which is a bijective transformation. One can therefore use eqs.(\ref{eq:generic:H}) to (\ref{eq:generic:h}) above with $ \langle \ln F_1 \rangle_{X_1} = \mu_{X_1}$, 
$\sigma_{X_1}$  and $\Sigma_{X_1}^{(m)}$ being the mean, variance and $m$-dimensional covariance matrix of the Gaussian 
variable $X_1$~\cite{Zografos2005}.
The entropy $H(Y_{F_1}) $ can be computed using the mean $\mu_Y$ and variance $\sigma_
Y^2$ of $Y_{F_1}$: 
\begin{align}
\mu_{X_1}&=\ln\left(\frac{\mu_{Y}^2}{\sqrt{\sigma^2_{Y}+\mu_{Y}^2}}\right)\,,\,
\sigma_{X_1}=2\ln\left(1+\frac{\sigma_{Y}^2}{\mu_{Y}^2}\right).
\end{align}
Writing $I^{(m,p,\tau)}(Y_{F_1})$ thus only requires to express the correlation function of the underlying Gaussian process $X_1$ as a function of the targeted $c_Y(\tau)$: 
\begin{equation}\label{eq:cx'}
c_{X_1}(\tau)=\frac{1}{\sigma_{X_1}^{2}} \ln((e^{\sigma_{X_1}^2}-1)c_{Y}(\tau)+1). 
\end{equation}
This permits to compute $h^{(m,\tau)}$ and $I^{(m,p,\tau)}$ analytically, using eqs.(\ref{eq:generic:MI}) and (\ref{eq:generic:h}). 

For stationary log-normal process $Y_{F_2}$, $F_2$ is not bijective. The entropy $H$ can be computed analytically, but this is not the case for $I^{(m,p,\tau)}$ and $h^{(m,\tau)}$.

\subsubsection{Correlation structures: short-range versus long range dependencies}

To study the impact of the correlation structure, processes with short-range dependencies are compared against processes with long-range dependencies, the former being well-represented by an exponentially decaying covariance,
\begin{align}
c_{Y}(\tau) & = \sigma_{Y}^2 e^{-f_{\rm c} |\tau|}, \; \; f_{{\rm c}}>0,
\label{eq;cexp}
\end{align}
and the latter by an algebraically decaying covariance, 
\begin{align}
 c_{Y}(\tau)&=\frac{\sigma_{Y}^2}{2} \left[(\tau-1)^{2{\cal H}}-2\tau^{2{\cal H}}+(\tau+1)^{2{\cal H}}\right]  \,,
\label{eq;ctau}
\end{align}
with $0.5<{\cal H}<1$.

\subsection{Two-point dependencies: Auto mutual information versus correlation}

\begin{figure*}[h]
\begin{center}
\includegraphics[width=0.75\linewidth]{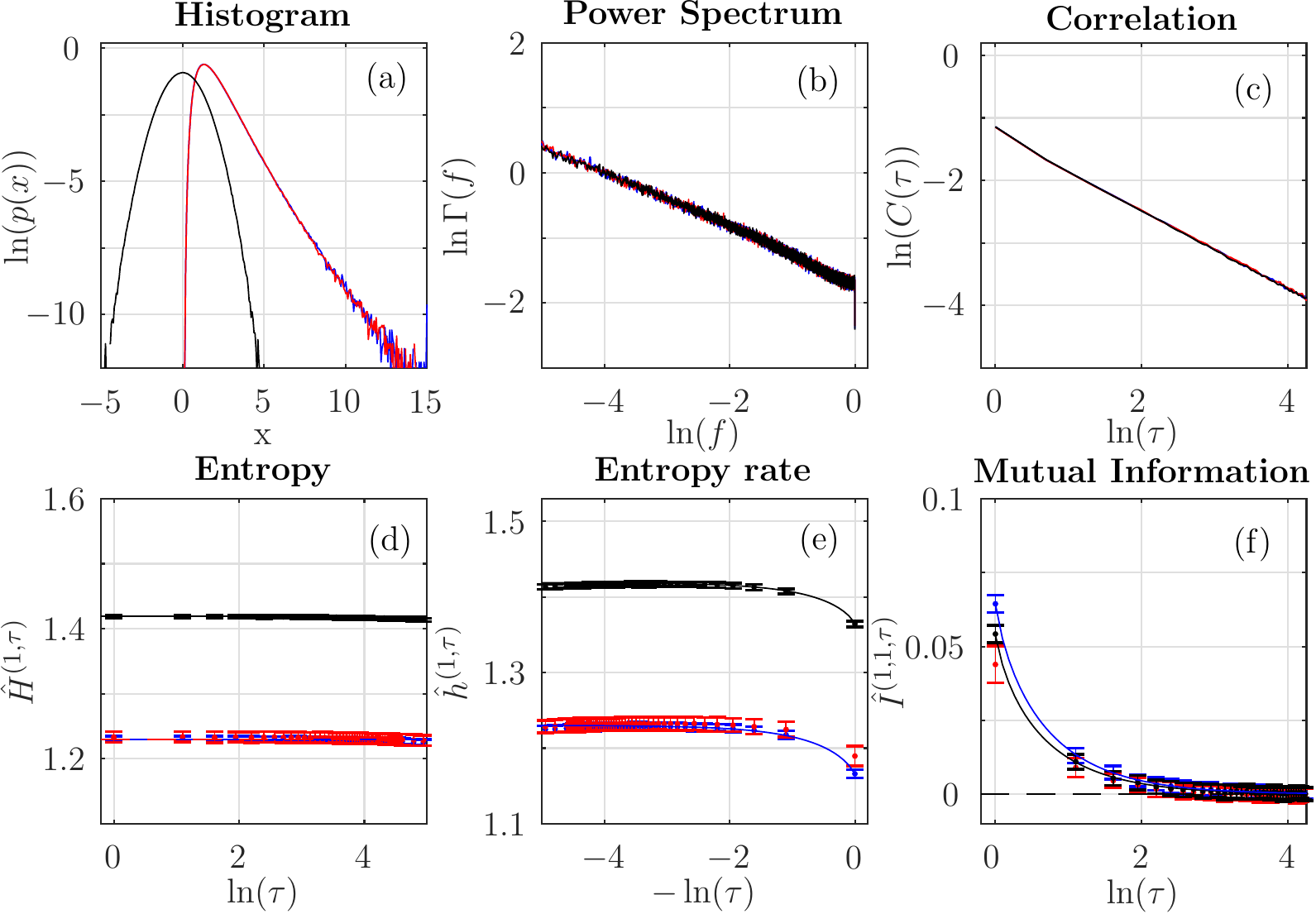}
\caption{{\bf Long-range dependencies. } Classical analysis: histogram (a), power spectrum (b) and correlation function (c). Information theory analysis for $m=1$, $p=1$: entropy (d), entropy rate (e) 
and auto mutual information (f). 
The color indicates the process: blue for standard log-normal ($Y_{F_1}$), red for $Y_{F_2}$ log-normal and black for Gaussian ($Y_{\mathbb I}$).
The continuous lines in (e,d,f) correspond to analytical expressions. 
 We used $N=2^{16}$ and $k=5$.}
\label{fig:results:expp92}
\end{center}
\end{figure*}

For both short-range and long range correlations, we compare the two LN processes obtained with $F_1$ and $F_2$, denoted $Y_{F_1}$,  $Y_{F_2}$, against a Gaussian process with same correlation function, denoted $Y_{{\mathbb I}}$, where ${\mathbb I}$ stands for the identity transform. 
First order statistics (probability density function) and second order statistics measured by the correlation function or the power spectral density as well as two point statistics ({\em i.e.}, $m=p=1$) entropy, entropy rate and AMI estimates are compared on Fig.~\ref{fig:results:expp9} for $Y_{{\mathbb I}}$, $Y_{F_1}$ and $Y_{F_2}$ with short-range dependences and on Fig.\ref{fig:results:expp92} for long-range dependences. 

As expected, for each pair of short range and long range processes, the two log-normal processes,  $Y_{F_1}$ and $Y_{F_2}$ cannot be distinguished one from the other using either PDF, correlation or power spectral estimation. 
Entropy being fully prescribed by the sole PDF, it is further not surprising that the two log-normal processes are also undistinguishable when using entropy estimates  $H$ (Fig~(\ref{fig:results:expp9},d)).

However, and interestingly, the estimates $I^{(1,1,\tau)}$ and $h^{(1,\tau)}$ of two-point AMI and entropy rate show differences as functions of $\tau$ between processes $Y_{F_1}$ and $Y_{F_2}$, both for the short-range and long-range correlation cases.
$I^{(1,1,\tau)}$  decreases and $h^{(1,\tau)}$ increases when $\tau \rightarrow \infty$, a predictable result as all dependencies vanish for large $\tau$ for stationary processes. Both functions (AMI and entropy rate) for processes $Y_{F_1}$ and $Y_{F_2}$ differ for small $\tau$, thus quantifying the difference between the joint distributions of these two processes, despite identical correlation functions and marginal distributions. 

Fig.~\ref{fig:results:expp9} and Fig.~\ref{fig:results:expp92} show that the two-point AMI estimates $\hat{I}^{(1,1,\tau)}$ for log-normal process $Y_{F_1}$ ressembles far more to that of the Gaussian process $Y_{\mathbb I}$ than to that of the companion log-normal process $Y_{F_2}$, even they all three have the same correlation function. 
Very interestingly, it provides  a deep insight into the temporal dynamics of the time series: 
processes $Y_{F_1}$ and $Y_{F_2}$ have same marginal distributions and same correlations functions, but the temporal dynamics of $Y_{F_1}$ is close to that of the Gaussian process $Y_{{\mathbb I}}$ while that of $Y_{F_2}$ is not.  
This also illustrates the benefits of AMI ${I}^{(1,1,\tau)}$ over entropy rate ${h}^{(1,\tau)}$ that mixes the impact of both marginal distribution and joint dynamical structure, thus making $Y_{F_2}$ comparable to $Y_{{\mathbb I}}$ at small $\tau$ when dependencies matter and $Y_{F_2}$ comparable to $Y_{F_1}$ at large $\tau$ when only the marginal distribution contribute to ${h}^{(1,\tau)}$. 

Figs~\ref{fig:results:expp9},d,e,f and \ref{fig:results:expp92},d,e,f  also report theoretical $H$, $h$ and AMI for processes $Y_{\mathbb I}$ and $Y_{F_1}$, while theoretical $h$ and AMI  are not available for process $Y_{F_2}$. 
All estimates well match the predicted theoretical behaviors (solid lines in Figs~\ref{fig:results:expp9}, \ref{fig:results:expp92}). The theoretical behaviors of AMI and entropy rate of process $Y_{F_1}$ are obtained using eqs.~(\ref{eq:generic:MI}) and (\ref{eq:generic:h}) with the analytical expression (\ref{eq:cx'}).

These observations essentially hold both for the short-range (Fig.~\ref{fig:results:expp9}) and long-range (Fig.~\ref{fig:results:expp92}) correlations and yield the first key conclusion of this work:
as opposed to correlation or spectral estimation, estimates of the two-points AMI (${I}^{(1,1,\tau)}$)  and entropy rate (${h}^{(1,\tau)}$) are able to quantify differences in the temporal dynamics of processes even when they share the exact same marginal distributions and correlation functions, as long as they have different two-point joint statistics. 

\section{Estimation performance and illustrations}

The aim of this section is to assess, by means of Monte Carlo simulations, the performance of the estimators $\hat{H}$, $\hat{I}^{(m,p,\tau)}$ and $\hat{h}^{(m,\tau)}$ for processes with different statistical structures and to study the impacts of key parameters either entering the definition of the estimators (number of nearest neighbors $k$) or characterizing the dependence structures of the processes (strength of the correlation $c_Y(\tau)$) as well as with sample sizes $N$.

%%%%%%%%%%%%%%%%%%%%%%%%
\subsection{Estimation performance}
\label{ssec:Procedure}

\subsubsection{Monte Carlo simulations Set-up}

The processes are synthetized numerically using circulant matrix embedding strategies, with the toolbox designed in \cite{Helgason2011,Helgason2011a} and made available at \href{http://www.hermir.org}{www.hermir.org}. As we have access to analytical curves only for $Y_{\mathbb{I}}$ and $Y_{F_1}$, these are the processes used to characterize the performance of the estimators.

For short and long memory, parameters are set $f_{\rm c}=0.1$ and ${\cal H}=0.7$ respectively.
All variances are set to $\sigma_Y \equiv 1$. 
For each case, $100$ realizations of processes are synthesized and performances are assessed as averages (and standard deviations) across realizations. 

To study the impact of the marginal distribution on estimation performance, process $Y_{F_1}$ is compared to process $Y_{\mathbb{I}}$.

\subsubsection{Dependence with the sample size $N$}

\begin{figure}[t]
\begin{center}
\includegraphics[width=.49\textwidth]{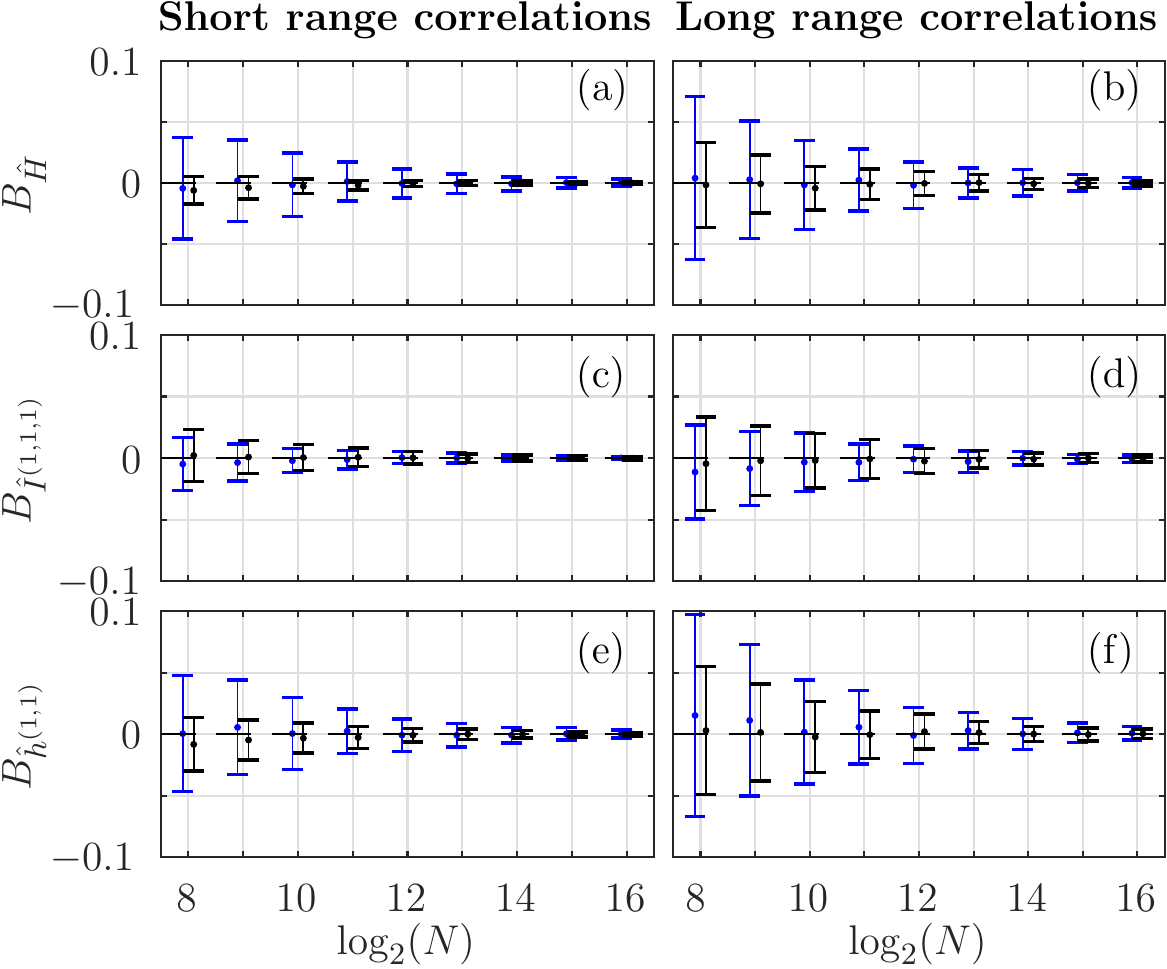}
\caption{Estimation performance for  $\hat{H}$ (entropy), $\hat{I}^{(1,1,\tau)}$ (auto mutual information) and $\hat{h}^{(1,\tau)}$ (entropy rate), as functions of $N$, for log-normal (blue) and Gaussian (black) processes, with short-range (left) and long range (right) correlations. 
$f_{\rm c}=0.1$,  $\mathcal{H}=0.7$, $c(\tau)=0.32$ and $k=5$. 
%The errorbars correspond to the standard deviation.
}
\label{fig:results:expInvvsN}
\end{center}
\end{figure}

Fig.~\ref{fig:results:expInvvsN} reports performances as functions of the sample size $N$, for $ 2^8 \leq N \leq 2^{16}$, for a fixed $k=5$. 
It clearly shows that the three estimators are asymptotically consistent (with vanishing biases and variances) irrespective of the marginal distribution and of the correlation being short-range or long-range. 
Unsurprisingly, variances are larger for the log-normal processes than they are for the Gaussian ones.
Also, long-range dependencies imply much larger variances compared to short-range dependencies for a same $c_Y(\tau)=0.32$ (which corresponds to $\tau=1$ for the long range correlation process and $\tau=11$ for the short range one), thus clearly showing that long memory does not induce biases but does imply larger variances. 
The proposed estimators are overall robust to marginal distributions and dependence structures~\cite{Grassberger2004,Gao2016}, and have satisfactory performance as soon as $N\geq2^{10}$, hence $N=2^{16}$ is used hereafter.

\subsubsection{Dependence with the number of nearest neighbors $k$} 

\begin{figure}[t]
\begin{center}
\includegraphics[width=.49\textwidth]{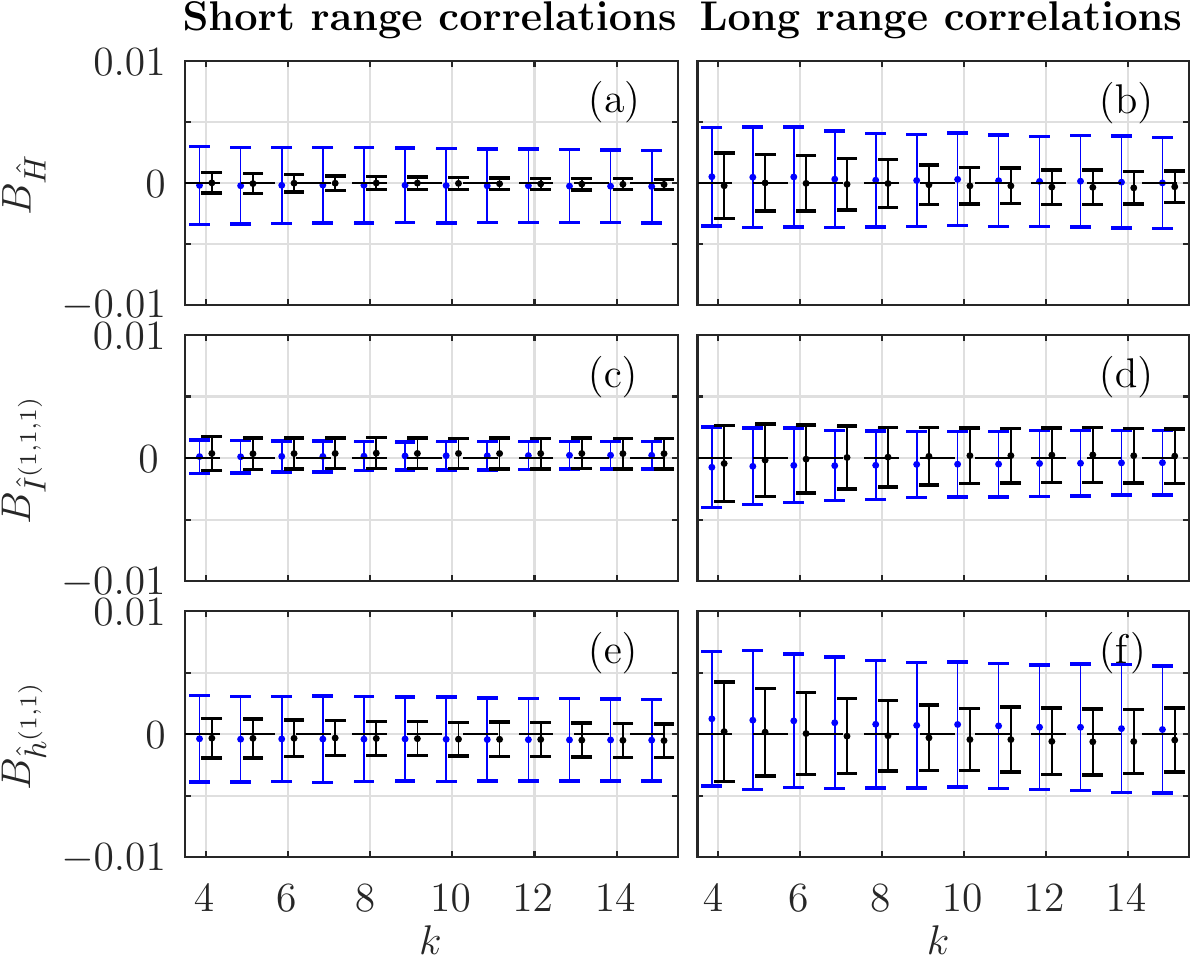}
\caption{
Estimation performance for  $\hat{H}$ (entropy), $\hat{I}^{(1,1,\tau)}$ (auto mutual information) and $\hat{h}^{(1,\tau)}$ (entropy rate), as functions of $k$, for log-normal (blue) and Gaussian (black) processes, with short-range (left) and long range (right) correlations. 
$f_{\rm c}=0.1$,  $\mathcal{H}=0.7$, $c(\tau)=0.32$ and $N=2^{16}$. }
\label{fig:results:expInvvsk}
\end{center}
\end{figure}

Fig.~\ref{fig:results:expInvvsk} reports estimation performance as functions of $k$, for $N=2^{16}$ and same correlation $c_Y=0.32$ as above. Performances barely depend on $k$, but for a mild increase of the biases, and this irrespective of the marginal distributions and of short-range or long-range nature of the correlations. 
Because increasing $k$ further implies increasing the computational load, $k=5$ is used hereafter.

\subsubsection{Influence of the correlation strength $c_Y(\tau)$} 
\label{sec:corr:strength}

\begin{figure}[t]
\begin{center}
\includegraphics[width=.49\textwidth]{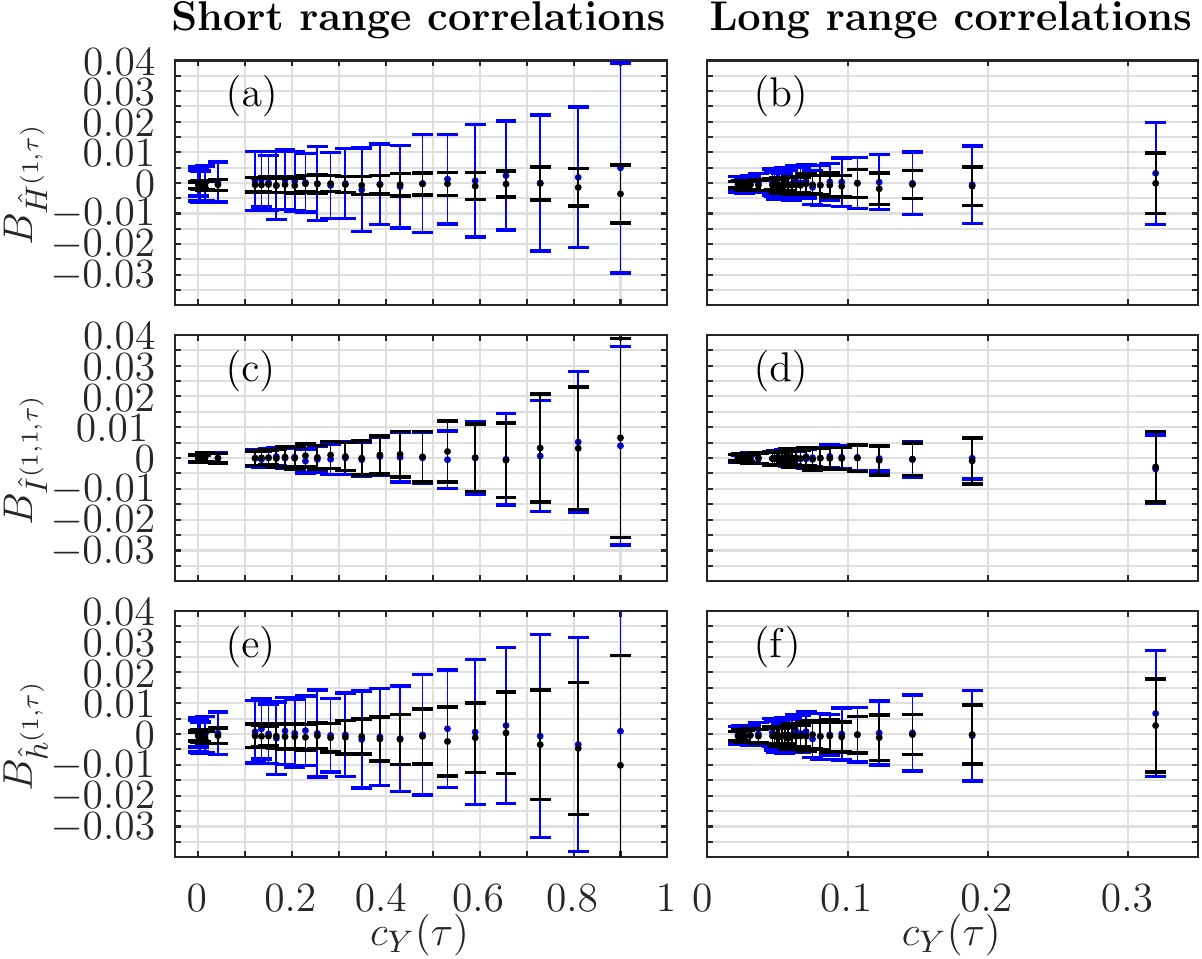}
\caption{
Estimation performance for $\hat{H}$ (entropy), $\hat{I}^{(1,1,\tau)}$ (auto mutual information) and $\hat{h}^{(1,\tau)}$ (entropy rate), as functions of the correlation strength $c_Y$, for log-normal (blue) and Gaussian (black) processes, with short-range (left) and long range (right) correlations. 
$f_{\rm c}=0.1$,  $\mathcal{H}=0.7$, $k=5$ and $N=2^{12}$. 
}
\label{fig:results:AMIvscorr}
\end{center}
\end{figure}

Fig.~\ref{fig:results:AMIvscorr} reports estimation performance as a function of $c_Y(\tau)$ and shows that 
all biases slightly increase when $c_Y(\tau)$ increases. The variance appears to increase a little faster than linearly in $c_Y$.
For both short-range and long-range correlations, the log-normal processes show larger standard deviations compared to the Gaussian ones, yet with similar biases. 
Further, for a given correlation level $c_Y(\tau)$ the long-range correlation structure shows larger standard deviations compared to  the short-range correlation. 

From this characterization, we can conclude that the entropy, AMI and entropy rate estimators show a low bias and standard deviation for the combination $k=5$ and $N=2^{16}$ of parameters used along the article. In addition, we can conclude that the standard deviation of all the three estimators is larger for strong correlations and non-Gaussian distributions, but their biases depend slightly on the marginal distribution  and the correlation strength.

%%%%%%%%%%%%%%%%%%%%%%%%%%%%%%%%%%%%%%%%%%%
\subsection{Process discrimination with auto mutual information}
\label{ssec:AMIvs2corr}

In Figs~\ref{fig:results:expp9},e,f and \ref{fig:results:expp92},e,f, we show that both AMI and entropy rate across scales are able to discriminate between processes $Y_{F_1}$ and  $Y_{F_2}$. The discrimination is far more effective for the smaller scales for both correlation structures (Long and short range).

\begin{figure}[t]
\begin{center}
\includegraphics[width=.49\textwidth]{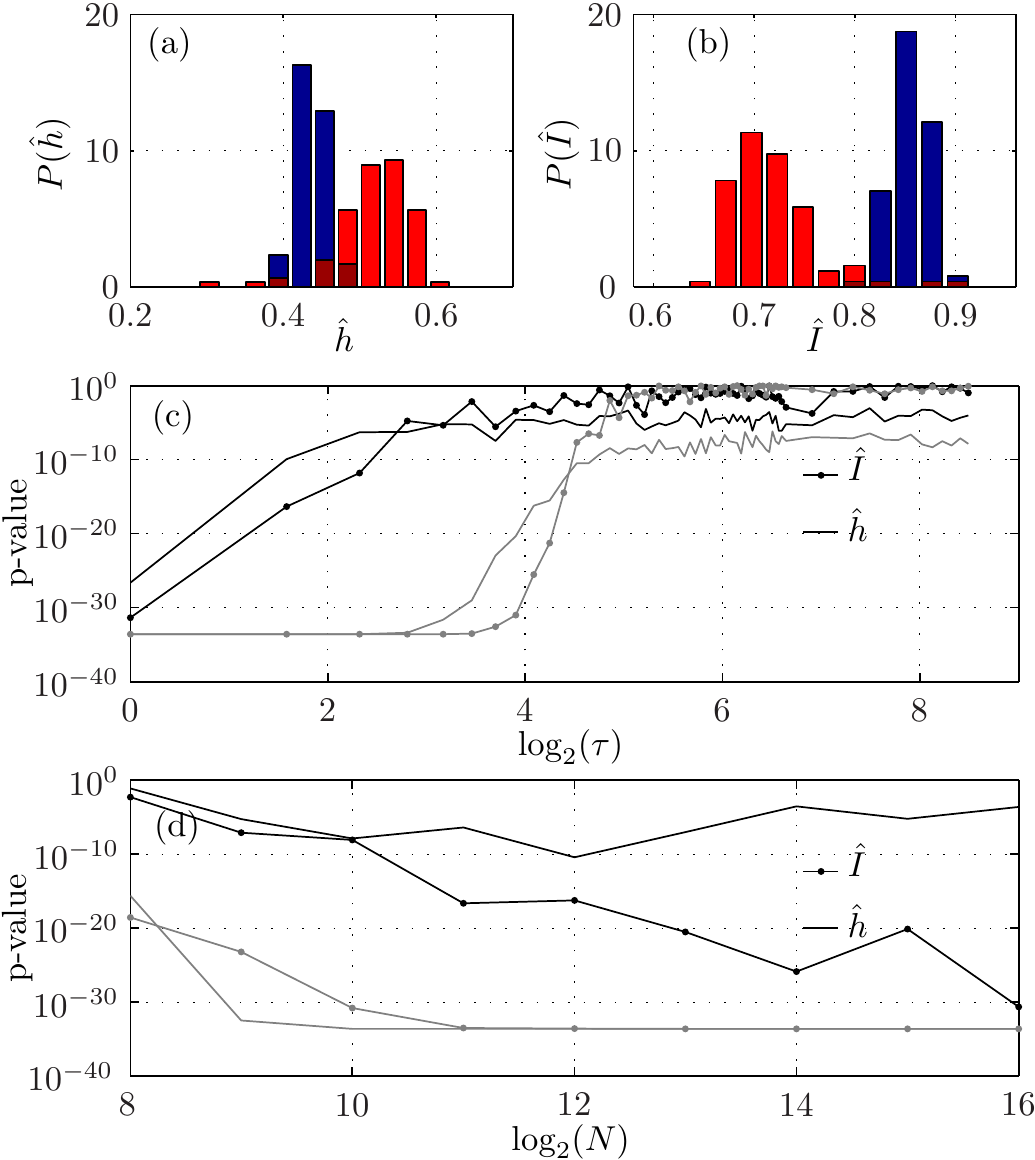}
\caption{{\bf Statistical test.} Histogram of the estimation of $\hat{h}^{(1,1)}(1)$ (a) and $\hat{I}^{(1,1)}(1)$ (b) for the scale invariant processes (blue for $Y_{F_1}$ and red for $Y_{F_2}$ of size $N=2^{16}$. 
(c) $p$-value of the Wilcoxon rank-sum test between the two log-nomal processes  $Y_{F_1}$ and $Y_{F_2}$
as fonction of the scale for the statistics of $\hat{h}^{(1,1)}(\tau)$ (dotted line) and $\hat{I}^{(1,1)}(\tau)$ (full line) with $N=2^{16}$.
(d) same quantities as (c) for scale $\tau=1$ as functions of the process size $N$. The statistics has been done on 100 realizations of both process. In (c) and (d) the gray curves correspond to the results obtained on the Exponential Decay processes. 
Mann-Whitney U-test ranksum.
}
\label{fig:results:pvalue}
\end{center}
\end{figure}

To quantify the differences between $Y_{F_1}$ and  $Y_{F_2}$ that we found in Figs~\ref{fig:results:expp9} and \ref{fig:results:expp92}, we first present the histograms of the estimations of the entropy rate (Fig.~\ref{fig:results:pvalue}a) and of the AMI (Fig.~\ref{fig:results:pvalue}b) for fixed $N=2^{16}$ and $\tau=1$. 
We perform a statistical ranksum test using Mann-Whitney U-test (Wilcoxon rank-sum test) and compute the $p$-value, for varying $\tau$ and $N$. 
As seen in Fig.~\ref{fig:results:pvalue}d, the $p$-value quickly decreases with increasing $N$, as expected. 
Fig.~\ref{fig:results:pvalue}c quantifies the distinguishability between both log-normal processes across scales, showing that the discrimination is far better for smaller scales. This confirms the observations of Figs~\ref{fig:results:expp9},e,f and \ref{fig:results:expp92},e,f.

\subsection{Dependence structure beyond correlations} 

This sections further explores the ability of AMI ${I}^{(m,p,\tau)}$, with $m \geq 1 $ and/or $p\geq 1$ to probe multiple-point dependences and hence fine details in temporal dynamics. 

\begin{figure}[t]
\begin{center}
\includegraphics[width=.49\textwidth]{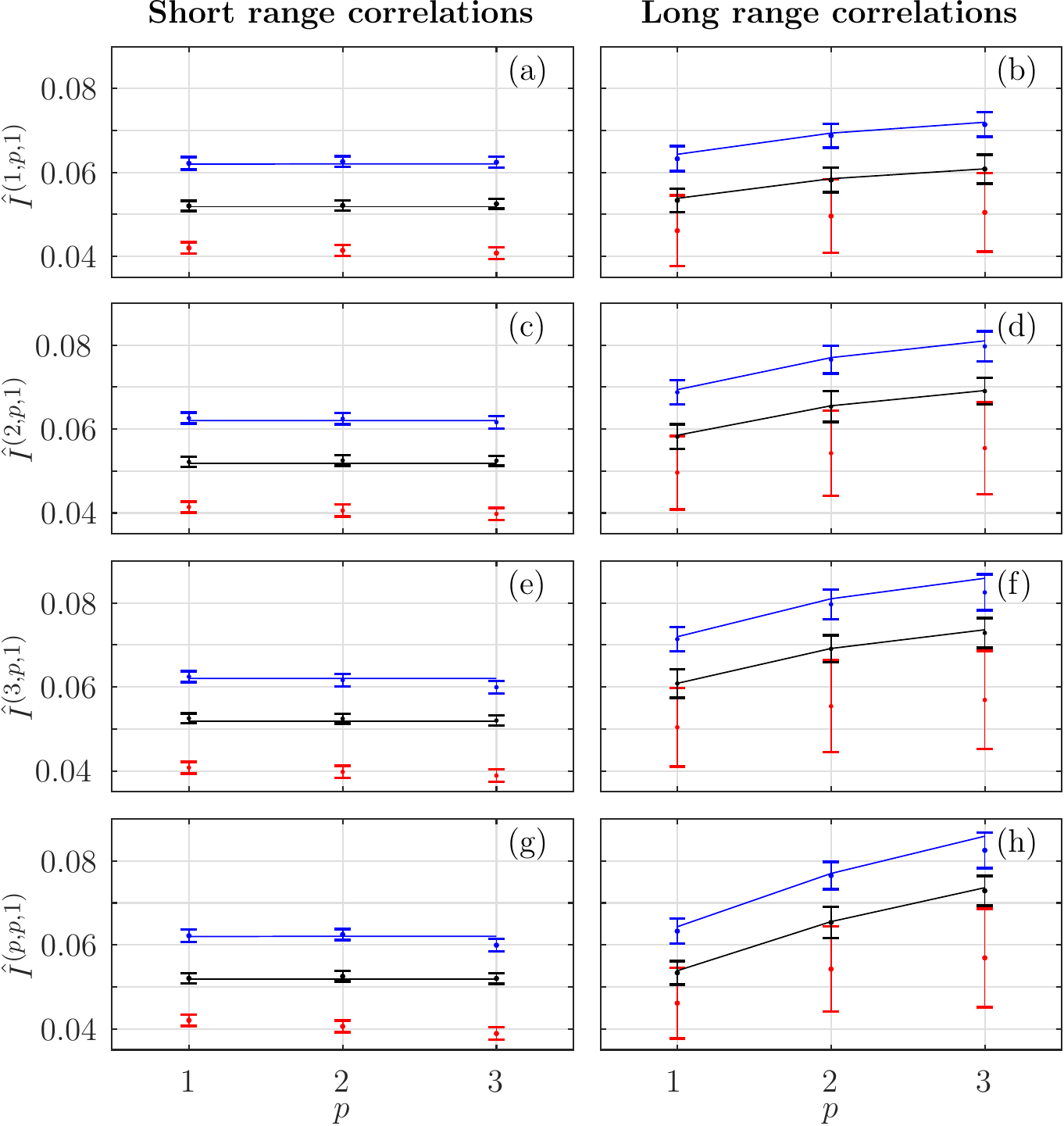}
\caption{{\bf Evolution of AMI with embedding.} 
$\hat{I}^{(m,p)}(\tau=1)$  for processes with exponentially decaying correlation (left column) and with power law decay (right column). Black is for $Y_{\mathbb I}$, blue is for $Y_{F_1}$ and red for  $Y_{F_2}$. 
First three lines are for fixed $m$ and increasing $p$: $m=1$ (a,b), $m=2$ (c,d) and $m=3$ (e,f). 
Last line is for $m=p$.
Other parameters are $N=2^{16}$, $k=5$, $\tau=1$.} 
\label{fig:results:embed}
\end{center}
\end{figure}

Fig.~\ref{fig:results:embed} reports higher-order AMI estimates $\hat{I}^{(m,p,\tau)}$ for several combination of $m$ and $p$, as functions of the embedding dimension $p$ for both short-range (left) and long-range (right) correlation functions, for processes $Y_{\mathbb I}$, $Y_{F_1}$  and $Y_{F_2}$. From top to bottom the $m$ value is equal to $1$, $2$, $3$ and $p$. Consequently, the first three lines study the behaviour of AMI on $p$ for fixed value of $m$, while the fourth line shows the behavior of AMI for $m=p$.
  
While two-point AMI yielded essentially identical conclusions for short-range and long-range correlation processes, higher order AMI reveals clear differences: 
For short-range correlation processes (left column), $Y_{\mathbb I}$, $Y_{F_1}$ and $Y_{F_2}$, $\hat{I}^{(m,p,\tau)}$ and ${I}^{(m,p,\tau)}$ when it can be computed analytically, show a very mild dependence of AMI with $m$ and $p$ for a fixed $\tau$.
To the converse, for long-range correlation processes (right column), $Y_{\mathbb I}$, $Y_{F_1}$ and $Y_{F_2}$, $\hat{I}^{(m,p,\tau)}$ and ${I}^{(m,p,\tau)}$ when it can be computed analytically, show constant increase of AMI when $p$ or $m$ increase, for a fixed $\tau$.
This fundamental observation clearly indicates that for short-range correlation, whatever $\tau$, there is little or no gain in the analysis of temporal dynamics brought by the study of higher order AMI. 
Conversely, for long-range correlation processes, ${I}^{(m,p,\tau)}$ captures more and more information in the temporal dynamics as the embedding dimensions are increased thus revealing a significant difference between short-range and long-range correlation structures.
Analytical formulae for ${I}^{(m,p,\tau)}$ when $m, p > 1$, are only available for the identity ${\mathbb I}$ and the natural transform $F_1$, namely for $Y_{{\mathbb{I}}}$ and $Y_{F_1}$.

\begin{figure}[t]
\begin{center}
\includegraphics[width=.49\textwidth]{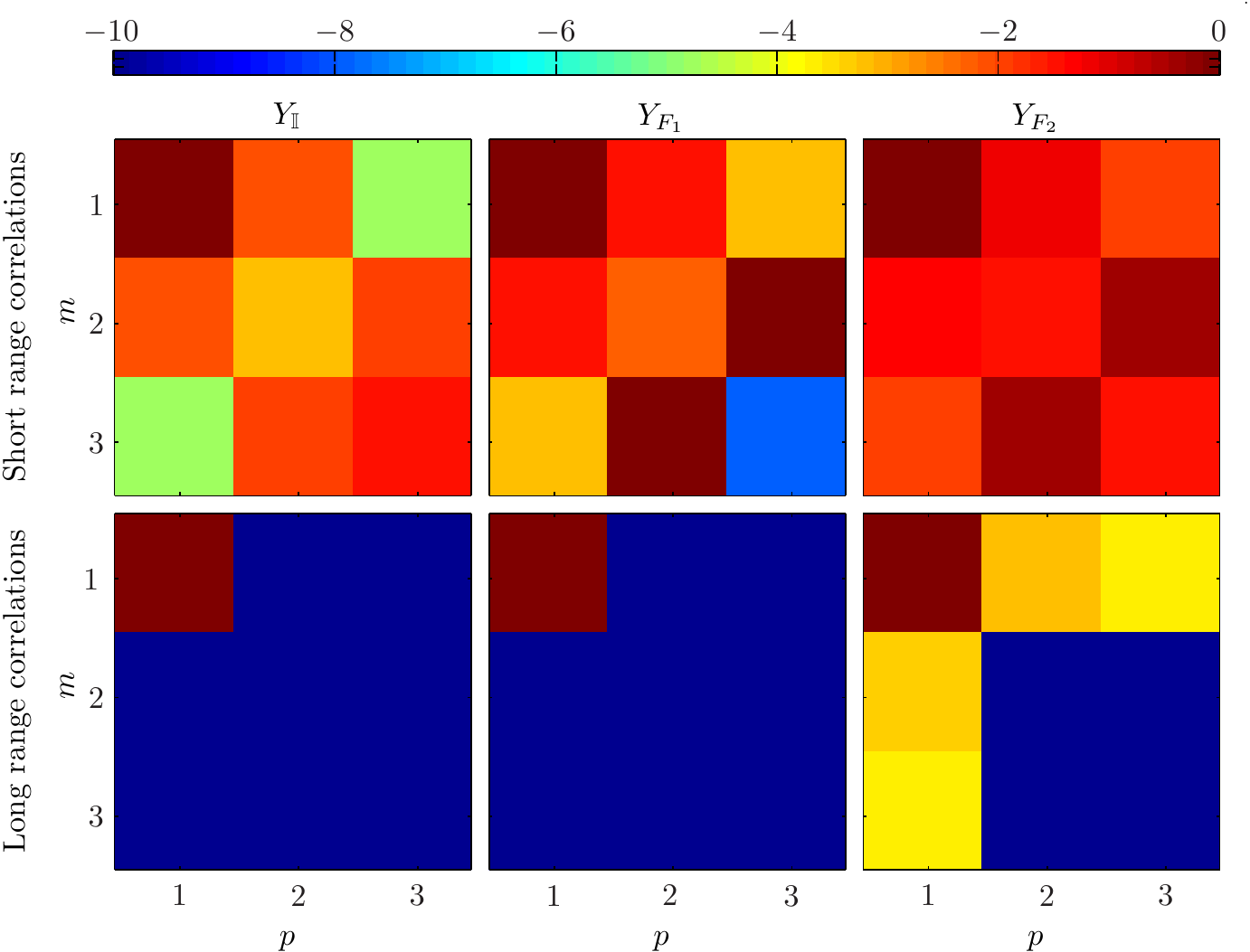}
\caption{{\bf Statistical test.} $p$-value of Wilcoxon rank-sum test between the statistics of $\hat{I}^{(1,1)}(1)$  and $\hat{I}^{(m,p)}(1)$  for processes with exponentially decaying correlation (top line) or with a power law decay (bottom line). Left column is for $Y_{\mathbb I}$, middle column for $Y_{F_1}$ and right column for $Y_{F_2}$.}
\label{fig:results:test2}
\end{center}
\end{figure}

In order to quantify the evolution of AMI when the embedding increases, we compute the $p$-value of Wilcoxon test between the statistic of $\hat{I}^{(1,1,1)}$  and $\hat{I}^{(m,p,1)}$ for the same processes as in Fig.~\ref{fig:results:embed}. Fig.~\ref{fig:results:test2} quantifies the increase of AMI when the embedding increases. This evolution of AMI is far more important in the long range correlations case, as expected from Fig.~\ref{fig:results:embed}. 
Further, table~\ref{tab:test} shows that, for fixed $m+p$, AMI evolves differently depending on the precise values of $m$ and $p$ separately. For a fixed $m+p$ AMI is larger when $m$ and $p$ are close, which can be understood as a closer similarity between the two embedded vectors appearing as the arguments of the AMI in eq.(\ref{eq:def:AMI}). 

\begin{table}[h]
\begin{center}
\begin{tabular}{||c||c|c|c||}\hline\hline
& $Y_{\mathbb{I}}$ & $Y_{F_1}$ & $Y_{F_2}$\\ \hline
Short range correlations & 0.34 &    0.96  &  0.72  \\ \hline
Long range correlations & $6e^{-15}$ &  $5e^{-20}$ & $2e^{-3}$\\ \hline
\end{tabular}
\caption{{\bf Statistical test.} $p$-value of Wilcoxon test between the statistics of   $\hat{I}^{(1,3)}(1)$  and $\hat{I}^{(2,2)}(1)$. } 
\label{tab:test}
\end{center}
\end{table}

%%%%%%%%%%%%%%%%%%%%%%%%%%%%%%%%%%%%%%%%%%%
\section{Conclusions and perspectives}
\label{ssec:DiscConc}

We explored the evolution of AMI ($I^{(m,p,\tau)}$) and entropy rate ($h^{(m,\tau)}$) on the scale parameter $\tau$. 
Even when considering only two point interactions ($m=p=1$) AMI and entropy rate probe statistics of any order, and hence non-linear dependences, and as such appear as unambiguous generalizations of the correlation function and the power spectrum. 

As an illustration, we analyzed two log-normal processes with identical 1-point distribution and identical correlation function.
AMI ($I^{(1,1,\tau)}$) and entropy rate ($h^{(1,\tau)}$) are able to discriminate these two processes, which appear as identical for classical analysis, by enlightening differences in high order dependences.
The behavior of these quantities along scales reveal the existence of stronger high-order dependences at smaller scale, which allows an easier separation of processes.

In addition, AMI generalizes easily to consider explicitly interactions between more than two points.
We thus probe the complexity of the dependence structure above linear order, {\em i.e.},
the additional information measured by AMI when considering an extra point should indicate the existence and relevance of next-order interactions.
As an example, comparing the effect of increasing the order of interactions for two dependence structures showed that while exponential decay ---~short range~--- dependences do not involve next order 
interactions, power law~--- long range~--- dependences do.

The same qualitative results were obtained when considering others one-point distribution, which supports the validity of the methodology for any distribution and dependence structure. This 
generality should make the methodology very interesting to perform non-Gaussian processes characterization in several different applications.

\bibliographystyle{IEEEtran}
\bibliography{biblio}

\end{document}